\documentclass[11pt,A4paper]{article}

\usepackage{amsthm}
\usepackage{amsmath}
\usepackage{natbib}
\usepackage[colorlinks,citecolor=blue,urlcolor=blue,filecolor=blue,backref=page]{hyperref}
\usepackage{graphicx}
\usepackage{adjustbox}
\usepackage{bm}
\usepackage{pdflscape}
\usepackage{afterpage}
\usepackage{a4wide}

\title{Uncertainty representation for early phase clinical test evaluations: a case study}

\author{Sara Graziadio$^{1}$ and Kevin J. Wilson$^{2}$ \\
$^{1}$NIHR Newcastle In Vitro Diagnostics Co-operative Newcastle, \\ Newcastle upon Tyne Hospitals NHS Foundation Trust, UK \\
$^{2}$School of Mathematics, Statistics \& Physics, Newcastle University, UK}
\date{}

\begin{document}

\maketitle

\begin{abstract}
In early clinical test evaluations the potential benefits of the introduction of a new technology into the healthcare system are assessed in the challenging situation of limited available empirical data. The aim of these evaluations is to provide additional evidence for the decision maker, who is typically a funder or the company developing the test, to evaluate which technologies should progress to the next stage of evaluation. In this paper we consider the evaluation of a diagnostic test for patients suffering from Chronic Obstructive Pulmonary Disease (COPD). We describe the use of graphical models, prior elicitation and uncertainty analysis to provide the required evidence to allow the test to progress to the next stage of evaluation. We specifically discuss inferring an influence diagram from a care pathway and conducting an elicitation exercise to allow specification of prior distributions over all model parameters. We describe the uncertainty analysis, via Monte Carlo simulation, which allowed us to demonstrate that the potential value of the test was robust to uncertainties. This paper provides a case study illustrating how a careful Bayesian analysis can be used to enhance early clinical test evaluations.
\end{abstract}

{\bf Keywords:}
Bayesian network;
Care pathway analysis;
Expert elicitation;
Sensitivity analysis;
Uncertainty analysis

\section{Introduction}

Clinical tests include in vitro tests and medical devices used by healthcare professionals or patients to support the diagnosis, screening or monitoring of a disease or disease stage. Tests need to be fully evaluated to assess their benefits and harms to patients and to the healthcare system before they can be introduced. Evaluations of clinical tests should assess not only the clinical and economic benefits, but also the potential change in the clinical decision making process if the test is adopted \citep{Kip18, Gra20}. In the early stages of evaluation, uncertainty is relatively high and little or no empirical data are available. In this context, decision analysis, structured prior elicitation and uncertainty analysis within a Bayesian framework, can provide defensible evidence on the potential benefits and disbenefits of a clinical test to decision makers.

Clear visualisation of the decision making process may support the test developers and funders in identifying the role of the test which could maximise the potential benefits of the test in the healthcare system. Identifying the role of the test is tightly linked to understanding where the clinical test could fit within the current care pathway. Care pathways are schematic visualisations of the journey of a patient in the healthcare system, and care pathway analysis is the comparison between the current care pathway and the proposed care pathway upon adoption of the new technology \citep{Cha19,Bra20}. In health economic evaluations they are used as preliminary models, established in collaboration with clinicians and other stakeholders, from which the economic model is extrapolated \citep{Bai15,Bri06}. However, methods for robust inference and analysis of care pathways are still lacking \citep{Chi10}, and the use of formal statistical approaches such as graphical models (in this case, influence diagrams) and structured expert judgement elicitation in early stage test evaluation is still relatively immature, although prior elicitation is an established approach within medical trials more generally \citep{Cha01,Tha19,Ped18}. An example of the use of influence diagrams within medical decision making is given in \citet{Owe97}. 

In this case study, the test of interest is a home monitoring test developed by a company for patients with Chronic Obstructive Pulmonary Disease (COPD). COPD is a chronic respiratory condition common in smokers and ex-smokers. There is no cure for patients affected by COPD but patients can be trained to recognise their symptoms and when they worsen. Worsening of symptoms can lead to exacerbations that are very unpleasant for the patient and can often lead to death. COPD is mainly monitored by patients at home but exacerbations often necessitate emergency care or consultation with a General Practitioner (GP). Sometimes patients manage their condition using steroids or antibiotics, but, when a patient suspects they are having an exacerbation, action can vary, with some patients presenting at hospital and others allowing symptoms to develop before seeking medical help. We hypothesised that by identifying exacerbations early the number of hospital visits could be reduced and effective treatment could be administered in the community.  

A company has developed a diagnostic test for home monitoring of COPD exacerbations. It is a urine test that monitors the level of urine biomarkers whose concentrations change when a patient moves from a stable condition to an exacerbation state. This test could aid patients in the self-management of their condition. 
We assumed weekly testing for the monitoring of their condition. The proposed use of the test is that patients would test themselves when they feel unwell and are concerned about their clinical symptoms. The test has the potential to identify whether patients are likely to experience an exacerbation. In this case, the patients will be prompted to take appropriate action, such as to take steroids or antibiotics or seek further medical help.

In this paper we consider the early evaluation of the test and its potential to impact the UK health system. 
We consider the initial care pathway developed for the test and detail how this was used to infer an influence diagram. We discuss the formal elicitation process used to obtain the quantities needed to populate the influence diagram. We describe how we use these quantities to solve the influence diagram and give an indication of the potential benefits of the test in the UK healthcare system. We consider the impact of uncertainties on the evaluation process. We discuss how the elicited quantities were used to form prior distributions on the model parameters and how they were used to assess the uncertainty on the key quantities in the analysis. We consider the sensitivity of the main model outputs to the uncertain model inputs. 

Throughout this case study we aim to illustrate the value of cutting-edge methodological developments in Bayesian analysis, to aid in uncertainty representation and decision support, in the context of a real practical problem. We conduct the elicitations using the state-of-the-art SHeffield ELicitation Framework \citep{Gos18,Oak16,Oha19}. The individual quantitative elicitations are of probabilities which must sum to one. This has recieved much attention in the recent literature \citep{Wil18,Elf16,Zap14,Elf13}. We specify conditional beta prior distributions for the probabilities as suggested in \citet{Wil18,Elf16}. The translation of qualitative statements into formal graphical models such as Bayesian networks, in this paper the translation of the care pathway to an influence diagram, is an important area of current research. Recent work in this area includes \citet{Wil18b,Smi10,Kol09}. Here we adapt the approach given in \citet{Smi10}. 

\section{Test evaluation}\label{s2} 

\subsection{Elicitation of the care pathway}\label{cp}

We selected 14 health professionals in the field of respiratory healthcare as our experts to support the development of the care pathway structure, under the guidance of the Academic Health Sciences Network (AHSN). Experts were selected for their complimentary perspectives and experience in COPD management, fulfilling the roles of GP, practice nurse, community matron, respiratory clinician, nurse in a nursing home and service development manager from the British Lung Foundation, and all had suitable practical knowledge.

\begin{figure}[ht]
	\centering
	\includegraphics[width = \linewidth]{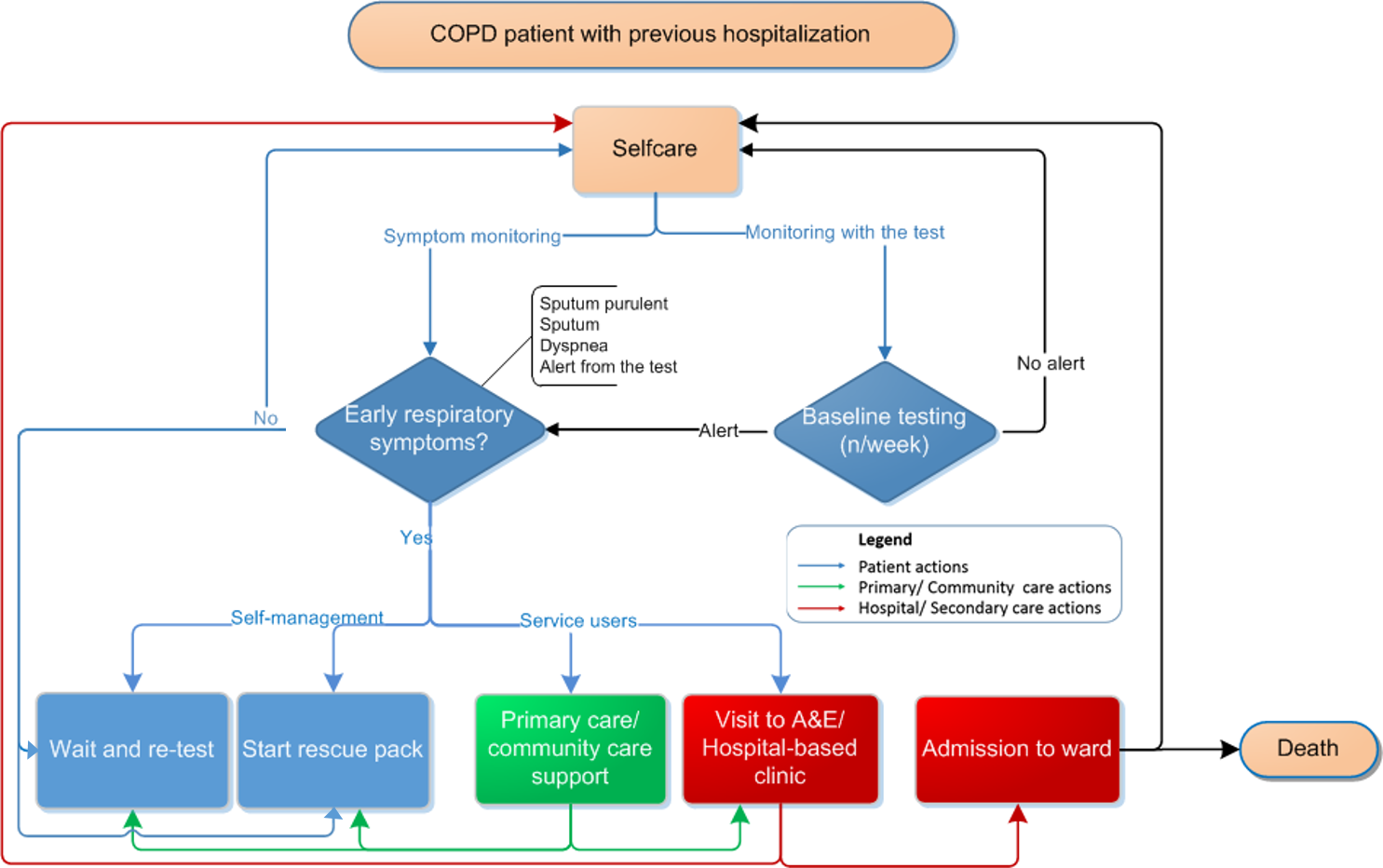}
	\caption{The final version of the care pathway for patients suffering for COPD. A\&E stands for Accident \& Emergency.}
	\label{fig:cp}
\end{figure}

The initial structure of the care pathway was inferred through a literature review and a review of guidelines for COPD management. The care pathway was discussed and its accuracy assessed by each interviewee through an iterative process until a consensus around the main structure of the pathway and the causal relationships between the nodes of the pathway was reached. In total, six iterations were used to establish the final care pathway. The resulting care pathway represents a high level picture of COPD management that could be considered common across a wide area of the UK. The final care pathway is given in Figure \ref{fig:cp}.

The care pathway shows, in the upper section, the monitoring loop of the patient, and, in the lower section, the possible actions of the patient, primary care professionals and secondary care professionals\footnote{In the UK, National Health Service (NHS) care is provided in two main ways: primary care (GPs and community services) and secondary care (hospitals and specialists).}. In the monitoring loop the patient monitors both the symptoms and the test. An alert can be triggered by a change of symptoms or a positive test result. If only a positive test result occurs associated with no change of symptoms during baseline testing, then the test result is not acted on and a re-test is advised. That is, the aim of this early modelling is to evaluate the test as a diagnostic tool, and excludes monitoring use of the test.

The lower part of the care pathway bifurcates and decisions of different healthcare professionals are represented in different colours: patient, primary care and secondary care decisions. With a suspected exacerbation, the patient could decide to wait and potentially retest, to treat their symptoms or to become service users by contacting their GP, a nurse or presenting at Accident \& Emergency (A\&E). The primary care professionals could decide to keep the patient at home and not treat them, to treat them or to send the patient to a hospital. Upon physical examination of the patient and in the presence of other diagnostic test results, the secondary care professionals are able to establish a correct diagnosis of whether an exacerbation is present. As a consequence, they may discharge the patient or admit them to a ward in the hospital.

\subsection{Inferring the influence diagram}\label{id}

Influence diagrams are graphical models which support the visualisation and solution of decision problems. They are an extension of Bayesian networks which incorporate decisions and utilities in addition to uncertainties. For comprehensive explanations of influence diagrams see \citet{Smi10,How81}. 

The first step to infer an influence diagram from the care pathway is the identification of the influential variables for the decision of the patients and clinicians involved in the care pathway. The care pathway facilitated the identification of the problem variables, which were the decisions of the patient, the primary care professionals and the secondary care professionals.

We used the approach of \citet{Smi10}, where the most influential variables were identified working backwards from the outcomes of potential decisions, and gradually identifying the different layers of uncertainties. We asked the experts to rank the variables to identify those which were most influential, to be kept in the analysis. There is some evidence that experts are more reliable in providing rankings than making direct judgements about unknowns \citep{Gos14}. The rankings from the different groups are omitted but can be found in \citet{Gra17}.

\begin{figure}[ht]
	\centering
	\includegraphics[width=\linewidth]{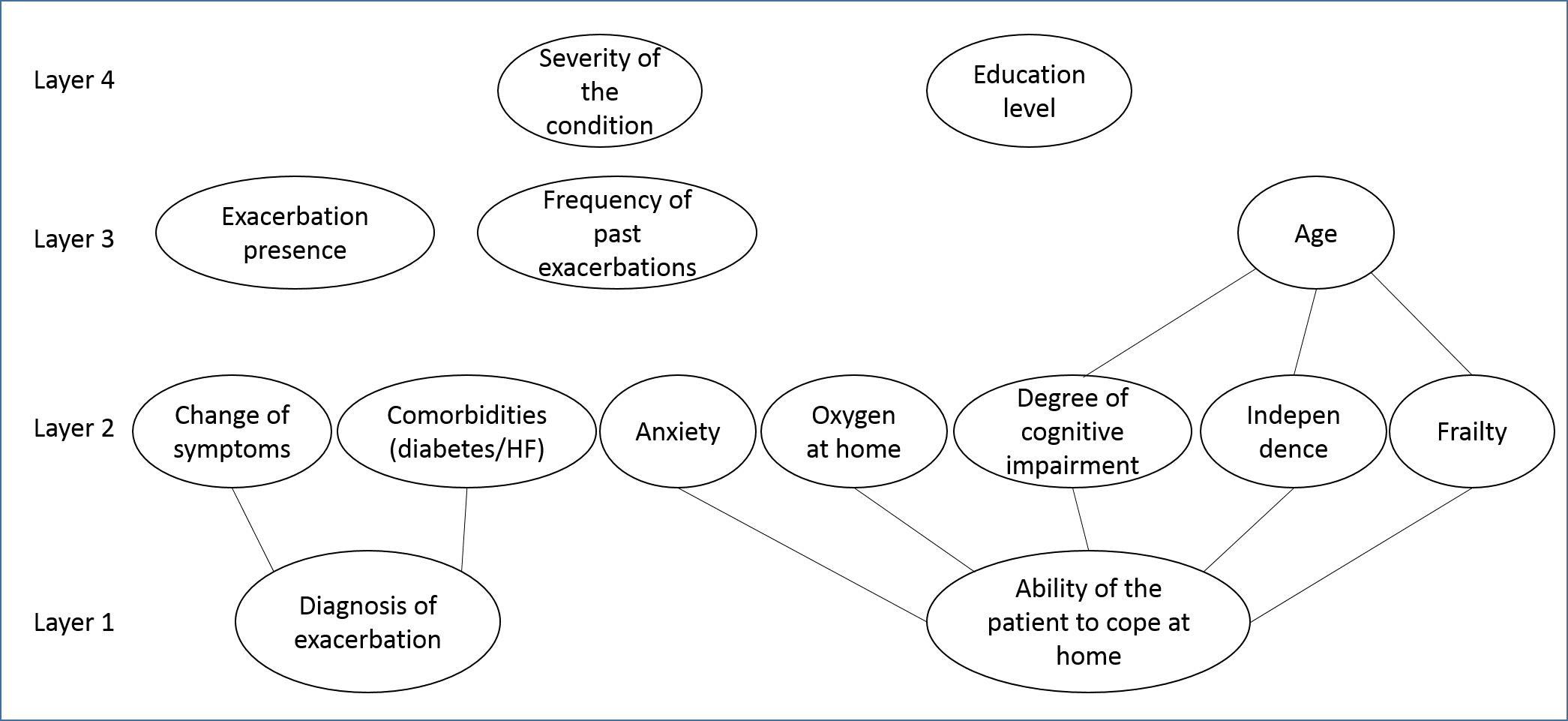}
	\caption{A trace back graph identifying the variables which are relevant to the decision making process for the management of COPD patients.}
	\label{trace}
\end{figure}

The rankings of the variables helped us to define the layers of uncertainty in the problem, which can be represented using the trace back graph \citep{Smi10} given in Figure \ref{trace}. In the trace back graph, connections between variables do not indicate dependency,  but indicate a partial ordering of how the variables influence each other.

From the discussions with experts, the first layer influencing the actions of the professionals and the patients was identified as the diagnosis. The experts also identified that their actions were also conditional on whether in their judgement the patient was able to cope with their current symptoms and broader condition (e.g. are they panicking? Do they live alone or do they have support?). However, it was only a minority of interviewees who agreed with this, and so it was disregarded in the further development of the influence diagram. The main elements affecting the decisions of the patients and healthcare professionals are elaborated in the second layer of the graph, and they were identified as the change of symptoms in the patient and their underlying comorbidities.

In the third layer the variables which were identified as being indirectly informative of a decision were included. Also, exacerbation presence is included in the third layer, as it cannot be observed directly by the clinician or the patient, but influences their diagnosis indirectly, mainly through the change in symptoms. In the fourth layer of uncertainty, the higher level information influencing the other layers included severity of condition and education level of the patient, which could influence the cognitive abilities of the patient. The elicitation process stopped at this point as later variables were not well defined and would not pass a clarity test \citep{Uff13}.

\begin{figure}[ht]
	\centering
	\includegraphics[height=3in]{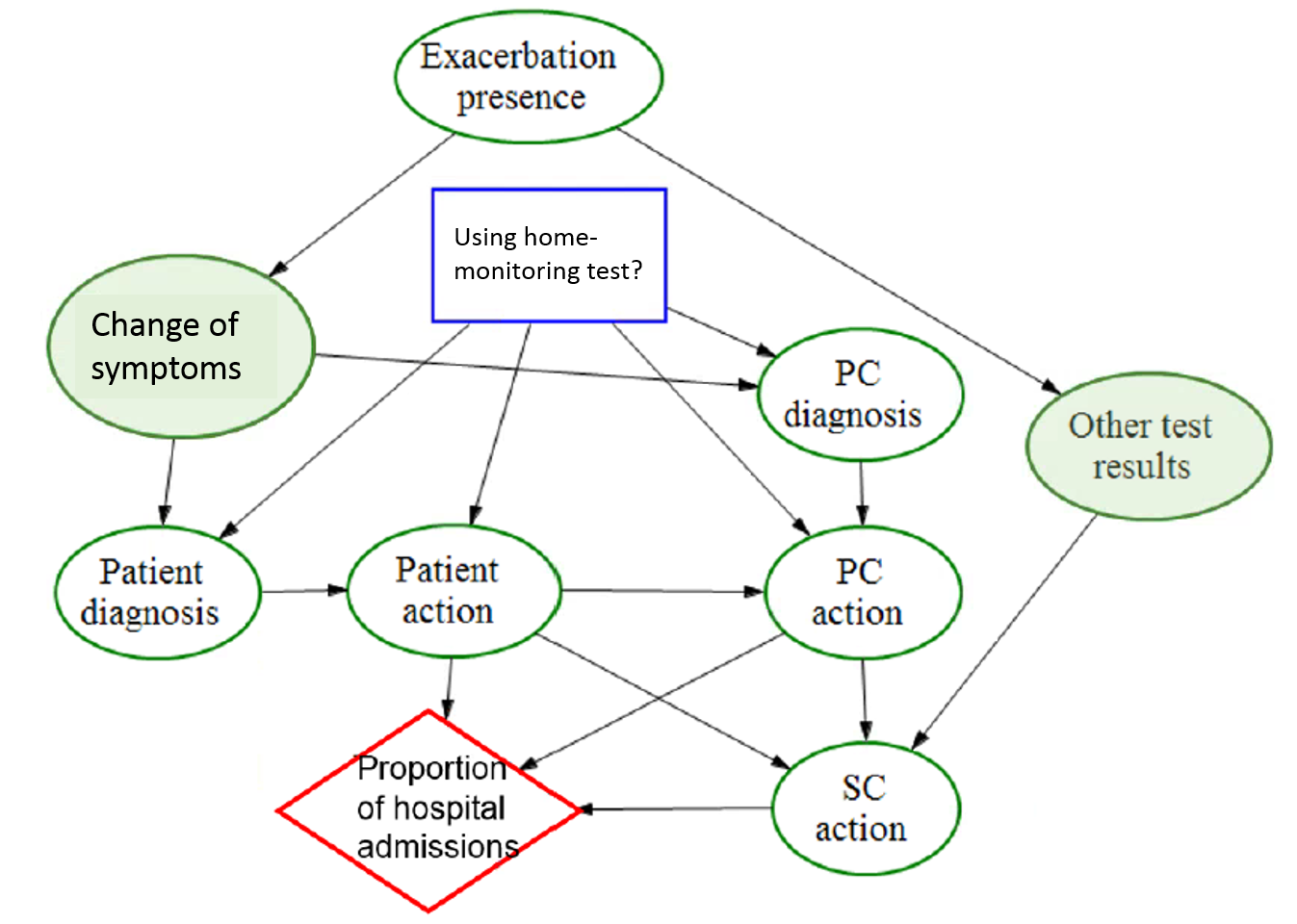}
	\caption{The complete influence diagram. Green circles are chance nodes, the blue rectangle is a decision node and the red diamond is a value node. PC is primary care and SC is secondary care.}
	\label{fig:id}
\end{figure}

The influence diagram (ID) was built using the collated judgements of the experts described above. The entry point of the ID is a patient experiencing an increase in symptoms which causes them concern. The full influence diagram is given in Figure \ref{fig:id}.

The decision node is the use of the test, as the decision maker is the funder (or the test developer). This node influences the diagnosis and actions of the patient and primary care clinicians, but not the actions of the secondary clinicians. The diagnosis is influenced by the change of symptoms. Exacerbation presence is not directly observed, and so is a chance node. The value node is the proportion of hospital admissions. The aim is to evaluate whether the introduction of the test into the current pathway could reduce the number of hospital admissions and visits, without increasing the risk of patient harm.

In this model, patient management is a high level representation. Only the final choice of the patient is modelled. The level of detail of this model has been calibrated to the uncertainty in the current evidence. The Bayesian framework is an advantage here, as the model can be updated and iterated in the next phase of test evaluation when more evidence is available.

Some of the factors which influence the decisions of patients and healthcare professionals were, at this early stage of test evaluation, unquantifiable. We used arc reversal techniques to reduce the ID to a model in which all variables are quantifiable \citep{Sha88}. The nodes that we need to remove, which are shaded green in Figure \ref{fig:id}, are not barren, as they have children, and so they need to be made barren before they can be removed.

\begin{figure}[ht]
	\centering
	\hbox{\includegraphics[height=6in]{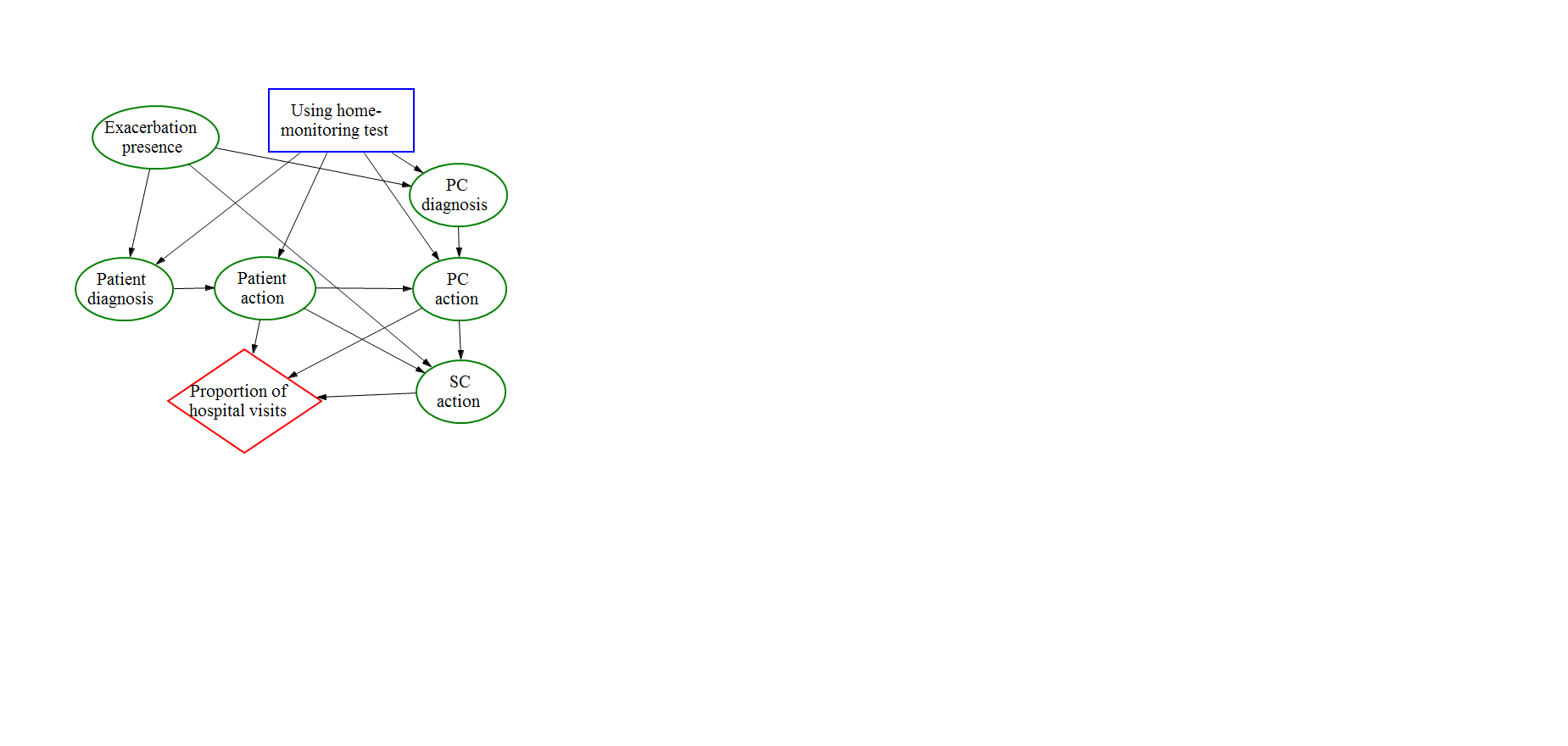}}
	\vspace{-1.5in}
	\caption{The reduced influence diagram. Green circles are chance nodes, the blue rectangle is a decision node and the red diamond is a value node. PC is prmary care and SC is secondary care.}
	\label{fig:id2}
\end{figure}

The easiest node to remove is ``Other test results'', as it has only one parent and one child. Using arc reversal, the edge between ``Other test results'' and ``SC actions'' can be reversed, making the node barren and so it can be removed. ``Exacerbation presence'' is a parent of ``Other test results'', and so when we remove it ``SC action'' must inherit it as a parent. To remove ``Change of symptoms'' we follow the same procedure to make it barren, adding an edge between ``Exacerbation presence'' and ``PC diagnosis'', and then between ``Exacerbation presence'' and ``Patient diagnosis'', to preserve the conditional independence structure. No cycles were created and independence properties between remaining variables were not altered. The resulting ID is given in Figure \ref{fig:id2}.

\subsection{Elicitation of expert judgements}\label{el}

In order to assess the possible effect of the introduction of the test on the number of hospital admissions requires populating the influence diagram with probabilities. In light of the scarcity of empirical data discussed earlier, this can instead be achieved using expert judgement elicitation. 

We used the SHELF method of behavioural expert judgement elicitation and aggregation \citep{Oak16,Oha06} as it offers a robust approach, suitable for early health technology assessment \citep{Gos14}. Within the SHELF framework a considerable amount of preparatory work preceded a workshop with the experts.

{\bf Identification of the quantities of interest}: The quantities of interest (QoIs) were the probabilities required to populate the ID. As a result of the edges in the ID, many of these were conditional probabilities. Some are constrained to 0 or 1. They are all given in Table \ref{fig:cond}. We also chose to elicit lower and upper 5\% quantiles for these probabilities to enable a comprehensive uncertainty analysis. The quantities of interest were test adoption ($X_0 \in $\{Yes, No\}), the presence of an exacerbation ($X_1 \in $\{Yes, No\}), patient diagnosis ($X_2 \in $\{Exacerbation, No Exacerbation\}), patient action ($X_3 \in $\{No Treatment, Treatment, Primary Care (PC), Secondary Care (SC)\}), primary care diagnosis ($X_4 \in $\{Exacerbation, No Exacerbation\}), primary care action ($X_5 \in $\{No Treatment, Treatment, Secondary Care (SC)\}) and secondary care action ($X_6 \in $\{Discharge, Admit\}).

\afterpage{%
	\clearpage
	\thispagestyle{empty}
	\begin{landscape}
		\begin{table}[ht]
			\vspace{-0.25in}
			\hspace{-0.5in}
			\scriptsize{
				\begin{tabular}{|c|c|c|c|c|c|c|c|}\hline
					\multicolumn{8}{|c|}{Decision node $X_0$} \\ \hline
					\multicolumn{4}{|l|}{} & $X_0=0$ & $X_0=1$ & \multicolumn{2}{c|}{} \\
					\multicolumn{4}{|l|}{Using test?}  & Yes & No & \multicolumn{2}{c|}{} \\ \hline
					\multicolumn{8}{|c|}{Prior probabilities for $X_1$} \\ \hline
					\multicolumn{4}{|l|}{} & $\Pr(X_1=0)$ & $\Pr(X_1=1)$ & \multicolumn{2}{c|}{} \\
					\multicolumn{4}{|l|}{$X_1$: Exacerbation presence}  & Yes & No & \multicolumn{2}{c|}{} \\ \hline
					\multicolumn{8}{|c|}{Conditional probabilities for $X_2$} \\ \hline
					\multicolumn{4}{|l|}{} & $\Pr(X_2=0\mid\bm X^{'})$ & $\Pr(X_2=1\mid\bm X^{'})$ & \multicolumn{2}{c|}{} \\ 
					\multicolumn{4}{|l|}{$X_2$: Patient diagnosis} & Exacerbation & No Exacerbation & \multicolumn{2}{c|}{} \\ \hline
					\multicolumn{2}{|l|}{} & Test: Yes & Exacerbation: Yes & Sensitivity & & \multicolumn{2}{c|}{} \\
					\multicolumn{2}{|l|}{} & Test: Yes & Exacerbation: No &  & Specificity & \multicolumn{2}{c|}{} \\
					\multicolumn{2}{|l|}{} & Test: No & Exacerbation: Yes &  & & \multicolumn{2}{c|}{} \\
					\multicolumn{2}{|l|}{$\bm X^{'}=(X_0,X_1)$} & Test: No & Exacerbation: No &  & & \multicolumn{2}{c|}{} \\ \hline
					\multicolumn{8}{|c|}{Conditional probabilities for $X_3$} \\ \hline
					\multicolumn{4}{|l|}{} & $\Pr(X_3=0\mid\bm X^{''})$ & $\Pr(X_3=1\mid\bm X^{''})$ & $\Pr(X_3=2\mid\bm X^{''})$ & $\Pr(X_3=3\mid\bm X^{''})$ \\ 
					\multicolumn{4}{|l|}{$X_3$: Patient action} & No treatment & Treatment & PC & SC \\ \hline
					\multicolumn{2}{|l|}{} & Test: Yes & Patient: Exacerbation & 0 & & & \\
					\multicolumn{2}{|l|}{} & Test: Yes & Patient: No Exacerbation & 1 & 0 & 0 & 0 \\
					\multicolumn{2}{|l|}{} & Test: No & Patient: Exacerbation & 0 & & & \\
					\multicolumn{2}{|l|}{$\bm X^{''}=(X_0,X_2)$} & Test: No & Patient: No Exacerbation & 1 & 0 & 0 & 0 \\ \hline
					\multicolumn{8}{|c|}{Conditional probabilities for $X_4$} \\ \hline
					\multicolumn{4}{|l|}{} & $\Pr(X_4=0\mid\bm X^{'})$ & $\Pr(X_4=1\mid\bm X^{'})$ & \multicolumn{2}{c|}{} \\ 
					\multicolumn{4}{|l|}{$X_4$: PC diagnosis} & Exacerbation & No Exacerbation & \multicolumn{2}{c|}{} \\ \hline
					\multicolumn{2}{|l|}{} & Test: Yes & Exacerbation: Yes &  & & \multicolumn{2}{c|}{} \\
					\multicolumn{2}{|l|}{} & Test: Yes & Exacerbation: No &  &  & \multicolumn{2}{c|}{} \\
					\multicolumn{2}{|l|}{} & Test: No & Exacerbation: Yes &  & & \multicolumn{2}{c|}{} \\
					\multicolumn{2}{|l|}{$\bm X^{'}=(X_0,X_1)$} & Test: No & Exacerbation: No &  & & \multicolumn{2}{c|}{} \\ \hline
					\multicolumn{8}{|c|}{Conditional probabilities for $X_5$} \\ \hline
					\multicolumn{4}{|l|}{} & $\Pr(X_5=0\mid\bm X^{'''})$ & $\Pr(X_5=1\mid\bm X^{'''})$ & $\Pr(X_5=2\mid\bm X^{'''})$ & \\ 
					\multicolumn{4}{|l|}{$X_5$: PC action} & No treatment & Treatment & SC & \\ \hline
					& Test: Yes & Patient Action: PC & PC: Exacerbation & 0 & &&  \\
					& Test: Yes & Patient Action: PC & PC: No Exacerbation & 1 & 0 & 0 & \\
					& Test: No & Patient Action: PC & PC: Exacerbation & 0 & & &   \\
					$\bm X^{'''}=(X_0,X_3,X_4)$ & Test: No & Patient Action: PC & PC: No Exacerbation & 1 & 0 & 0 & \\ \hline
					\multicolumn{8}{|c|}{Conditional probabilities for $X_6$} \\ \hline
					\multicolumn{4}{|l|}{} & $\Pr(X_6=0\mid\bm X^{''''})$ & $\Pr(X_6=1\mid\bm X^{''''})$ & \multicolumn{2}{c|}{} \\ 
					\multicolumn{4}{|l|}{$X_6$: SC action} & Discharge & Admit & \multicolumn{2}{c|}{}\\ \hline
					& Exacerbation: Yes & PC Action: SC & Patient Action: PC &  &   & \multicolumn{2}{c|}{} \\
					& Exacerbation: No & PC Action: SC & Patient Action: PC & 1 & 0 & \multicolumn{2}{c|}{} \\
					& Exacerbation: Yes & PC Action: Null & Patient Action: SC &  &  & \multicolumn{2}{c|}{} \\
					$\bm X^{''''}=(X_1,X_3,X_5)$ & Exacerbation: No & PC Action: Null & Patient Action: SC & 1 & 0 & \multicolumn{2}{c|}{} \\ \hline
			\end{tabular}}
			\caption{The conditional probability tables of the reduced ID. PC stands for primary care and SC stands for secondary care.}
			\label{fig:cond}
		\end{table}
	\end{landscape}
	\clearpage
}

{\bf Preparation of the evidence dossier:} The evidence dossier summarises all quantitative information relevant to the quantities of interest, and is compiled by the elicitation team and the experts. In this case there were reports of investigations into COPD in the literature, data from the test development provided by the company and data on COPD from national databases. Summaries of the information from each of these sources were reported to the experts in the elicitation.

{\bf Identification of the elicitation team}: The client for this project is the company. The roles of co-ordinator, who organises the workshop, and analyst, who displays the resulting probability distributions to the experts, were fulfilled by one author and the role of advisor, who helps to clear up misunderstandings between analyst and experts, by the other. The two authors shared the roles of facilitator, who manages the workshop and recorder, who takes notes (and an audio recording in this case).

{\bf Identification and recruitment of experts:} Typically between 4 and 8 experts are chosen to provide a diversity of experience and knowledge. We chose two experts from those used to construct the ID to take part in the probability elicitation: the Respiratory Programme Lead for the Academic Health Society Network and a GP, Clinical Advisor for Newcastle Hospitals and Senior Partner in a Surgery in Newcastle. All relevant data were shared between the analyst and the two experts.

{\bf Experts' contribution to the evidence dossier:} The experts contributed details of the Newcastle primary care co-operative COPD audit 2014/15 and data from the quality and outcomes framework, which collates the information needed for the national reward and incentives programme for GP surgeries in England. The GP also provided the numbers of patients in their surgery with COPD and the proportion with additional comorbidities and mental health issues.

There were then several stages in the workshop itself.

{\bf Training of the experts:} The experts were trained in subjective probabilities, the basic structure of influence diagrams and conditional probabilities. The structure of the ID was validated by the experts. The evidence dossier was discussed in a meeting prior to the workshop, and the technical training was provided on the day of the elicitation by the facilitator.

{\bf The elicitation session:} The main elicitation session was followed by a second meeting to validate the final ID including probabilities and uncertainty distributions. During the elicitation, for each QoI, the experts were asked to provide a lower quantile below the median, the median and an upper quantile above the median. For the lower and upper quantiles they were asked

``What is the value for the QoI that you would judge it unlikely that the true value is less/greater than this value?''

For the median they were asked

``What is your best estimate of the QoI?''.

This order was designed to avoid the experts anchoring on the median when choosing their upper and lower quantiles. The individual judgements were compared and if there was disagreement the method of the Rational Imaprtial Observer was used to reach a consensus \citep{Oak16}. That is, rather than giving their own judgements, they were asked to agree on the judgement of a rational observer who is aware of all of the information and has heard all of the discussions between the experts.

There were two probabilities that the experts were not comfortable giving their probabilities for; the proportion of patients admitted by secondary care clinicians in A\&E after self-referral and after referral from primary care. These quantities were subsequently elicited from two secondary care respiratory consultants. The resulting medians (``best estimate'') and upper and lower quantiles from the elicitation are given in Table \ref{fig:cond2}.

\afterpage{%
	\clearpage
	\thispagestyle{empty}
	\begin{landscape}
		\begin{table}[ht]
			\vspace{-0.25in}
			\hspace{-0.5in}
			\scriptsize{
				\begin{tabular}{|c|c|c|c|c|c|c|c|}\hline
					\multicolumn{8}{|c|}{Decision node $X_0$} \\ \hline
					\multicolumn{4}{|l|}{} & $X_0=0$ & $X_0=1$ & \multicolumn{2}{c|}{} \\
					\multicolumn{4}{|l|}{Using test?}  & Yes & No & \multicolumn{2}{c|}{} \\ \hline
					\multicolumn{8}{|c|}{Prior probabilities for $X_1$} \\ \hline
					\multicolumn{4}{|l|}{} & $\Pr(X_1=0)$ & $\Pr(X_1=1)$ & \multicolumn{2}{c|}{} \\
					\multicolumn{4}{|l|}{}  & Yes & No & \multicolumn{2}{c|}{} \\ 
					\multicolumn{4}{|l|}{$X_1$: Exacerbation presence} & 0.45 (0.4-0.55) & 0.55 (0.5-0.6) & \multicolumn{2}{c|}{} \\ \hline
					\multicolumn{8}{|c|}{Conditional probabilities for $X_2$} \\ \hline
					\multicolumn{4}{|l|}{} & $\Pr(X_2=0\mid\bm X^{'})$ & $\Pr(X_2=1\mid\bm X^{'})$ & \multicolumn{2}{c|}{} \\ 
					\multicolumn{4}{|l|}{$X_2$: Patient diagnosis} & Exacerbation & No Exacerbation & \multicolumn{2}{c|}{} \\ \hline
					\multicolumn{2}{|l|}{} & Test: Yes & Exacerbation: Yes & 0.9 (0.8-0.97) & 0.1 (0.03-0.2) & \multicolumn{2}{c|}{} \\
					\multicolumn{2}{|l|}{} & Test: Yes & Exacerbation: No & 0.1 (0.03-0.3) & 0.9 (0.7-0.97) & \multicolumn{2}{c|}{} \\
					\multicolumn{2}{|l|}{} & Test: No & Exacerbation: Yes & 0.7 (0.6-0.75) & 0.3 (0.25-0.4) & \multicolumn{2}{c|}{} \\
					\multicolumn{2}{|l|}{$\bm X^{'}=(X_0,X_1)$} & Test: No & Exacerbation: No & 0.3 (0.25-0.4) & 0.7 (0.6-0.75) & \multicolumn{2}{c|}{} \\ \hline
					\multicolumn{8}{|c|}{Conditional probabilities for $X_3$} \\ \hline
					\multicolumn{4}{|l|}{} & $\Pr(X_3=0\mid\bm X^{''})$ & $\Pr(X_3=1\mid\bm X^{''})$ & $\Pr(X_3=2\mid\bm X^{''})$ & $\Pr(X_3=3\mid\bm X^{''})$ \\ 
					\multicolumn{4}{|l|}{$X_3$: Patient action} & No treatment & Treatment & PC & SC \\ \hline
					\multicolumn{2}{|l|}{} & Test: Yes & Patient: Exacerbation & 0 & 0.35 (0.25-0.5) & 0.6 (0.4-0.8) & 0.05 (0.02-0.15) \\
					\multicolumn{2}{|l|}{} & Test: Yes & Patient: No Exacerbation & 1 & 0 & 0 & 0 \\
					\multicolumn{2}{|l|}{} & Test: No & Patient: Exacerbation & 0 & 0.15 (0.1-0.2) & 0.7 (0.5-0.8) & 0.15 (0.1-0.2) \\
					\multicolumn{2}{|l|}{$\bm X^{''}=(X_0,X_2)$} & Test: No & Patient: No Exacerbation & 1 & 0 & 0 & 0 \\ \hline
					\multicolumn{8}{|c|}{Conditional probabilities for $X_4$} \\ \hline
					\multicolumn{4}{|l|}{} & $\Pr(X_4=0\mid\bm X^{'})$ & $\Pr(X_4=1\mid\bm X^{'})$ & \multicolumn{2}{c|}{} \\ 
					\multicolumn{4}{|l|}{$X_4$: PC diagnosis} & Exacerbation & No Exacerbation & \multicolumn{2}{c|}{} \\ \hline
					\multicolumn{2}{|l|}{} & Test: Yes & Exacerbation: Yes & 0.9 (0.8-0.97) & 0.1 (0.03-0.2) & \multicolumn{2}{c|}{} \\
					\multicolumn{2}{|l|}{} & Test: Yes & Exacerbation: No & 0.1 (0.03-0.2) & 0.9 (0.8-0.97)  & \multicolumn{2}{c|}{} \\
					\multicolumn{2}{|l|}{} & Test: No & Exacerbation: Yes & 0.7 (0.65-0.8)  & 0.3 (0.25-0.4) & \multicolumn{2}{c|}{} \\
					\multicolumn{2}{|l|}{$\bm X^{'}=(X_0,X_1)$} & Test: No & Exacerbation: No & 0.3 (0.25-0.5)  & 0.7 (0.5-0.75) & \multicolumn{2}{c|}{} \\ \hline
					\multicolumn{8}{|c|}{Conditional probabilities for $X_5$} \\ \hline
					\multicolumn{4}{|l|}{} & $\Pr(X_5=0\mid\bm X^{'''})$ & $\Pr(X_5=1\mid\bm X^{'''})$ & $\Pr(X_5=2\mid\bm X^{'''})$ & \\ 
					\multicolumn{4}{|l|}{$X_5$: PC action} & No treatment & Treatment & SC & \\ \hline
					& Test: Yes & Patient Action: PC & PC: Exacerbation & 0 & 0.8 (0.6-0.95) & 0.2 (0.05-0.4) &  \\
					& Test: Yes & Patient Action: PC & PC: No Exacerbation & 1 & 0 & 0 & \\
					& Test: No & Patient Action: PC & PC: Exacerbation & 0 & 0.85 (0.8-0.9) & 0.15 (0.1-0.2) &   \\
					$\bm X^{'''}=(X_0,X_3,X_4)$ & Test: No & Patient Action: PC & PC: No Exacerbation & 1 & 0 & 0 & \\ \hline
					\multicolumn{8}{|c|}{Conditional probabilities for $X_6$} \\ \hline
					\multicolumn{4}{|l|}{} & $\Pr(X_6=0\mid\bm X^{''''})$ & $\Pr(X_6=1\mid\bm X^{''''})$ & \multicolumn{2}{c|}{} \\ 
					\multicolumn{4}{|l|}{$X_6$: SC action} & Discharge & Admit & \multicolumn{2}{c|}{}\\ \hline
					& Exacerbation: Yes & PC Action: SC & Patient Action: PC & 0.3 (0.05-0.35) & 0.7 (0.65-0.95)  & \multicolumn{2}{c|}{} \\
					& Exacerbation: No & PC Action: SC & Patient Action: PC & 1 & 0 & \multicolumn{2}{c|}{} \\
					& Exacerbation: Yes & PC Action: Null & Patient Action: SC & 0.38 (0.05-0.45) & 0.62 (0.55-0.95) & \multicolumn{2}{c|}{} \\
					$\bm X^{''''}=(X_1,X_3,X_5)$ & Exacerbation: No & PC Action: Null & Patient Action: SC & 1 & 0 & \multicolumn{2}{c|}{} \\ \hline
			\end{tabular}}
			\caption{The complete conditional probability tables of the reduced ID. Lower and upper quantiles are given in brackets. PC stands for primary care and SC stands for secondary care.}
			\label{fig:cond2}
		\end{table}
	\end{landscape}
	\clearpage
}

The assumption that nobody without an exacerbation would be admitted by secondary care may seem strong. However, secondary care have access to a range of tests and carry out a more comprehensive assessment of a patient. Therefore, the diagnosis by the secondary care clinicians is regarded as a gold standard diagnosis.


\subsection{Solution to the influence diagram}\label{sol}

The aim of the study was to provide the decision maker with some early evidence about the potential benefits of the test. The test aims to reduce unnecessary hospital admissions, but not by increasing the risk of missing exacerbations which could lead to a worsening of patient conditions or death. We reduced the assessment of this aim to three queries:

{\bf Query 1} Would the proportion of hospital admissions for patients experiencing a true exacerbation be reduced, if the test was adopted? To answer this query required comparison of $\Pr(X_6=1\mid X_0=0)$, the probability of admission to hospital given that the test is adopted with $\Pr(X_6=1\mid X_0=1)$, the probability of admission to hospital given that the test is not adopted.

These two probabilities can be calculated based on the structure of the ID. They are
{\footnotesize
	\begin{eqnarray*}
		\Pr(X_6=1\mid X_0=0) & = & \Pr(X_6=1\mid X_1=0,X_3=2,X_5=2)\Pr(X_5=2\mid X_0=0,X_3=2,X_4=0) \\
		&&\times\Pr(X_4=0\mid X_0=0,X_1=0)\Pr(X_3=2\mid X_0=0,X_2=0) \\ &&\times\Pr(X_2=0\mid X_0=0,X_1=0)\Pr(X_1=0) \\
		&& + \Pr(X_6=1\mid X_1=0,X_3=3)\Pr(X_3=3\mid X_0=0,X_2=0) \\ &&\times\Pr(X_2=0\mid X_0=0,X_1=0)\Pr(X_1=0),
	\end{eqnarray*}
	\begin{eqnarray*}
		\Pr(X_6=1\mid X_0=1) & = & \Pr(X_6=1\mid X_1=0,X_3=2,X_5=2)\Pr(X_5=2\mid X_0=1,X_3=2,X_4=0) \\
		&&\times\Pr(X_4=0\mid X_0=1,X_1=0)\Pr(X_3=2\mid X_0=1,X_2=0) \\ &&\times\Pr(X_2=0\mid X_0=1,X_1=0)\Pr(X_1=0) \\
		&& + \Pr(X_6=1\mid X_1=0,X_3=3)\Pr(X_3=3\mid X_0=1,X_2=0) \\ &&\times\Pr(X_2=0\mid X_0=1,X_1=0)\Pr(X_1=0).
\end{eqnarray*}}
Using the probabilities reported in Table \ref{fig:cond2}, we obtain $\Pr(X_6=1\mid X_0=0)=0.046$ and $\Pr(X_6=1\mid X_0=1)=0.043$. This corresponds to a 6.5\% potential reduction in necessary hospital admissions after introduction of the test. 

{\bf Query 2} Would the proportion of missed exacerbations be reduced, if the test was adopted? The quantities which can be used to evaluate this are $\Pr(X_5=0, X_1=0\mid X_0=1) + \Pr(X_3=0, X_1=0\mid X_0=1)$, the probability that a patient is having an exacerbation and receives no treatment given that the test is not adopted, and  $\Pr(X_5=0, X_1=0\mid X_0=0) + \Pr(X_3=0, X_1=0\mid X_0=0)$, the probability that the patient is having an exacerbation and receives no treatment given that the test is adopted.

The probabilities of missed cases without and with the test are
{\footnotesize
	\begin{eqnarray*}
		\Pr(X_{3\textrm{ or } 5}=0, X_1=0\mid X_0=1) & = & \Pr(X_5=0\mid X_4=1,X_3=2,X_0=1)\Pr(X_4=1\mid X_0=1,X_1=0) \\
		&& \times \Pr(X_3=2\mid X_0=1,X_2=0)\Pr(X_2=0\mid X_0=1,X_1=0)\\ && \times\Pr(X_1=0) \\
		&& + \Pr(X_3=0\mid X_0=1,X_2=1)\Pr(X_2=1\mid X_0=1,X_1=0)\\ && \times\Pr(X_1=0),
	\end{eqnarray*}
	\begin{eqnarray*}
		\Pr(X_{3\textrm{ or } 5}=0, X_1=0 \mid X_0=0) & = & \Pr(X_5=0\mid X_4=1,X_3=2,X_0=0)\Pr(X_4=1\mid X_0=0,X_1=0) \\
		&& \times \Pr(X_3=2\mid X_0=0,X_2=0)\Pr(X_2=0\mid X_0=0,X_1=0) \\ && \times\Pr(X_1=0) \\
		&& + \Pr(X_3=0\mid X_0=0,X_2=1)\Pr(X_2=1\mid X_0=0,X_1=0) \\ && \times\Pr(X_1=0).
\end{eqnarray*}}
Using the probabilities reported in Table \ref{fig:cond2}, we obtain $\Pr(X_{3\textrm{ or } 5}=0, X_1=0\mid X_0=1)=0.201$ and $\Pr(X_{3\textrm{ or } 5}=0, X_1=0\mid X_0=0)=0.069$. Thus introduction of the test could reduce missed cases of true exacerbations by 66\%.

{\bf Query 3} Would the proportion of unnecessary visits to A\&E be reduced, if the test was adopted? The comparison in this case is between $\Pr(X_5=2, X_1=1\mid X_0=1)+\Pr(X_3=3,X_1=1\mid X_0=1)$, the probability that the patient is not having an exacerbation and goes to hospital conditional on no test and $\Pr(X_5=2, X_1=1\mid X_0=0)+\Pr(X_3=3,X_1=1\mid X_0=0)$, the probability that the patient is not exacerbating and goes to hospital conditional on test adoption.

The probabilities of unnecessary A\&E visits without and with the test are
{\footnotesize
	\begin{eqnarray*}
		\Pr(X_{3\textrm{ or }5}=\textrm{SC}, X_1=1\mid X_0=1) & = & \Pr(X_5=2\mid X_0=1,X_3=2,X_4=0)\Pr(X_4=0\mid X_0=1,X_1=1) \\
		&& \times \Pr(X_3=2\mid X_0=1,X_2=0)\Pr(X_2=0\mid X_0=1,X_1=1) \\ && \times\Pr(X_1=1) \\
		&& +\Pr(X_3=3\mid X_0=1,X_2=0)\Pr(X_2=0\mid X_0=1,X_1=1) \\ && \times\Pr(X_1=1),
	\end{eqnarray*}
	\begin{eqnarray*}
		\Pr(X_{3\textrm{ or }5}=\textrm{SC},X_1=1\mid X_0=0) & = & \Pr(X_5=2\mid X_0=0,X_3=2,X_4=0)\Pr(X_4=0\mid X_0=0,X_1=1) \\
		&& \times \Pr(X_3=2\mid X_0=0,X_2=0)\Pr(X_2=0\mid X_0=0,X_1=1) \\ && \times\Pr(X_1=1) \\
		&& +\Pr(X_3=3\mid X_0=0,X_2=0)\Pr(X_2=0\mid X_0=0,X_1=1)\\ && \times\Pr(X_1=1).
\end{eqnarray*}}
From the probabilities in Table \ref{fig:cond2}, the probabilities of interest are $\Pr(X_{3\textrm{ or }5}=\textrm{SC},X_1=1\mid X_0=1)=0.030$ and $\Pr(X_{3\textrm{ or }5}=\textrm{SC},X_1=1\mid X_0=0)=0.003$. This corresponds to a potential reduction of unnecessary A\&E visits of 90\% on adoption of the test.

In the next section, we consider the assessment of the uncertainty in these query probabilities which was conducted in the study.

\section{Uncertainty assessment}\label{3}

\subsection{Specification of uncertainty distributions}\label{pr}

We elicited from the experts three quantiles; the median, a quantile above the median and a quantile below the median. We take these latter quantiles to be the 5\% and 95\% quantiles. We chose to fit a beta distribution to each probability from the experts, as it is relatively flexible and has support on $[0,1]$. For events with more than two categories, for example $X_3$, a conditional beta distribution structure was used \citep{Wil18}. That is, beta distributions represented the probabilities that an individual belonged to particular categories conditional on their not belonging to any previous category.

Two methods to fit beta distributions from the three quantiles were considered; minimising the Euclidean distance between the quantiles and the beta distribution and an approach based on only two elicited quantities from \citet{Chr11}, which converts the mode and the 95\% quantile directly to values of the beta distribution parameters. The advantage of the latter method would be a shorter elicitation session, if reliability of the estimation could be demonstrated. 

Consider the first approach. Suppose that the three quantiles elicited from the expert were \\ $(q_e(0.05),q_e(0.5),q_e(0.95))$ and the equivalent quantiles for a beta distribution with parameters $(a,b)$ are $q_{a,b}(0.05),q_{a,b}(0.5),q_{a,b}(0.95)$. Then the parameters $(a,b)$ are chosen as
\begin{displaymath}
\textrm{argmin}_{a,b}\left\{[q_e(0.05)-q_{a,b}(0.05)]^2+[q_e(0.5)-q_{a,b}(0.5)]^2 +[q_e(0.95)-q_{a,b}(0.95)]^2\right\}.
\end{displaymath}

In the second approach, suppose that the elicited mode is $m$. Then the value of $a$, for any particular choice of $b$, can be found as
\begin{displaymath}
a = \dfrac{1+m(b-2)}{1-m},
\end{displaymath}
and the value of $b$ is chosen to match $q_e(0.95)$ in combination with this value of $a$.

We note that we regard the best estimate as the median and mode respectively in these approaches. While some of the distributions were relatively asymmetric, in most of the fitted distributions the mode and median were relatively close to eachother.

An alternative to the second approach would be to use the mode and 5\% quantile. We chose to use the 95\% quantile as we feel it is perhaps easier for an expert to think about the journey of ``most'' patients rather than a small proportion of patients.

The methods were compared based on the query probabilities from the previous section. Once each beta distribution on the input probabilities was defined using either method, a Monte Carlo simulation was conducted to obtain a sample of size 200,000 from each query probability. From this empirical distribution on each model output, we extracted the mean, median, mode, standard deviation and the 95\% implied probability interval. The probability intervals were calculated in two ways: using the empirical distribution from the Monte Carlo sample and assuming a beta distribution for the query probability with parameters $a=\bar{x}[\bar{x}(1-\bar{x})/s^2-1]$ and $b=(1-\bar{x})[\bar{x}(1-\bar{x})/s^2-1]$, where $(\bar{x},s)$ were the mean and standard deviation from the sample.

Table \ref{PI} provides the comparison based on the probability intervals.

\afterpage{%
\clearpage
\thispagestyle{empty}
\begin{landscape}
\begin{table}[ht]
	\centering {\small
		\begin{tabular}{|c|c|c|c|c|c|c|c|} \hline 
			\multicolumn{8}{|c|}{Query 1: Proportion of hospital admissions} \\ \hline
			& \multicolumn{2}{|l}{3-point estimates} &\multicolumn{2}{|l|}{2-point estimates} & & 3-point estimates &2-point 		estimates \\ \hline
			95\% PI & Beta & Empirical & Beta & Empirical & 90\% PI & Beta & Beta \\ \hline
			Test & (0.018, 0.082) & (0.019, 0.082) & (0.010, 0.104) & (0.011, 0.105) & Test & (0.021, 0.074) & (0.013, 0.092) \\
			No test & (0.034, 0.059) & (0.034, 0.060) & (0.019, 0.079) & (0.017, 0.077) & No test & (0.035, 0.057) & (0.022, 0.073) \\ \hline
			\multicolumn{8}{c}{}\\ \hline
			\multicolumn{8}{|c|}{Query 2: Proportion of missed exacerbations} \\ \hline
			& \multicolumn{2}{|l}{3-point estimates} &\multicolumn{2}{|l|}{2-point estimates} &  & 3-point estimates & 2-point estimates \\ \hline
			95\% PI & Beta & Empirical & Beta & Empirical & 90\% PI & Beta & Beta \\ \hline
			Test & (0.034, 0.125) & (0.034, 0.125) & (0.029, 0.138) & (0.030, 0.139) & Test & (0.038, 0115) & (0.034, 0.125) \\
			No test & (0.166, 0.239) & (0.167, 0.239) & (0.130, 0.287) & (0.131, 0.288) & No test & (0.172, 0.233) & (0.140, 0.272) \\ \hline 
			\multicolumn{8}{c}{}\\ \hline
			\multicolumn{8}{|c|}{Query 3: Proportion of unnecessary A\& E visits} \\ \hline
			& \multicolumn{2}{|l}{3-point estimates} &\multicolumn{2}{|l|}{2-point estimates} &  & 3-point estimates & 2-point estimates \\ \hline
			95\% PI & Beta & Empirical & Beta & Empirical & 90\% PI & Beta & Beta \\ \hline
			Test & (0.001, 0.010) & (0.001, 0.010) & (0.0001, 0.020) & (0.0001, 0.020) & Test & (0.001, 0.009) & (0.0002, 0.016) \\
			No test & (0.021, 0.041) & (0.021, 0.041) & (0.017, 0.048) & (0.017, 0.048) & No test & (0.022, 0.039) & (0.019, 0.044) \\ \hline 
	\end{tabular} }
	\caption{Comparison of implied probability intervals (labelled PI) obtained from the two- and three-point beta distribution estimates.}
	\label{PI}
\end{table}
\end{landscape}
\clearpage
}


There are differences between the resulting beta distributions when based on the two- or three-point estimates. We considered the difference between the median estimated from the simulation and that calculated exactly using the ID. The squared difference between these two quantities were $(0.001, 0.001, 0.000)$ for the three point estimate and $(0.003,0.001,0.0002)$ for the two-point estimate in the case of no test and $(0.000,0.000,0.0001)$ and $(0.002,0.001,0.0002)$ respectively in the case of test adoption. The values from the three-point estimate are consistently lower than the values from the two-point estimate. We see from the table that the distributions estimated from two quantiles have wider probability intervals than those estimated from three quantiles. The beta distribution assumption looks reasonable based on the similarity of the implied probability intervals with and without the assumption. 

In our assessment of the uncertainty in the query probabilities in the next section we will provide results based on both the two-point and three-point beta distribution estimates as we believe the uncertainty inherent in this estimation should be incorporated into the uncertainty analysis. We will focus the discussion around the three-point estimates as these displayed the best properties above.

\subsection{Assessing query uncertainty}\label{un}

\begin{figure}[ht]
	\centering
	\includegraphics[width=0.45\linewidth]{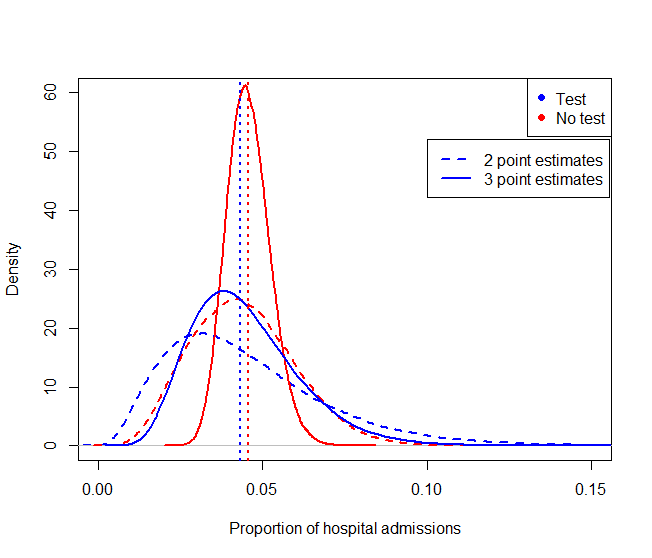}
	\includegraphics[width=0.45\linewidth]{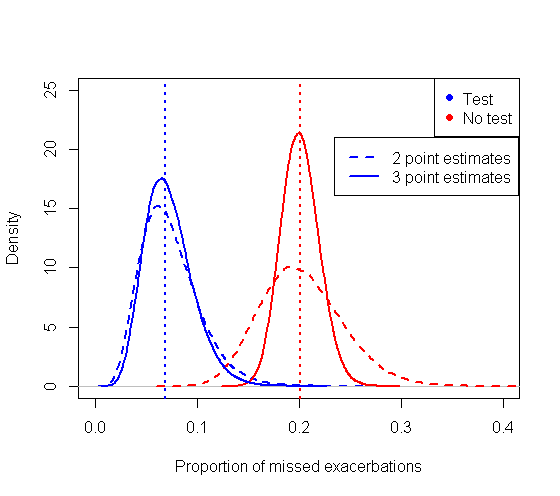}
	\includegraphics[width=0.4\linewidth]{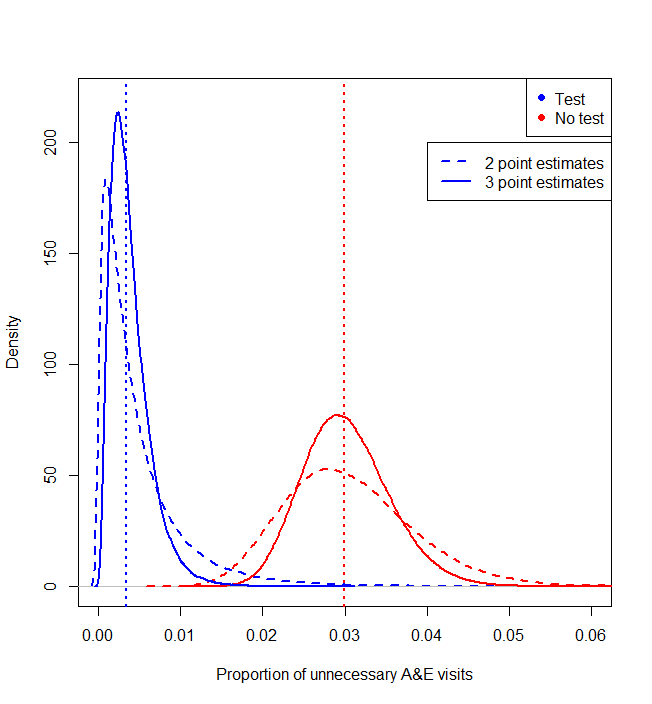}
	\caption{Densities of the uncertainty distributions for the queries comparing the current pathway with the pathway under test adoption. From the top left, clockwise we have query 1, query 2 and query 3.}
	\label{dens}
\end{figure}

{\bf Query 1} The top left plot in Figure \ref{dens} provides the densities of the beta distributions under the no test and test scenarios for the output quantity of query 1, the probability of hospital admission when a patient suspects an exacerbation. There is a large overlap between the distributions of the proportion of hospital admissions in the current care pathway and the pathway if the test was introduced. The distribution is more diffuse in the ``test'' than the ``no test'' condition as the experts were more confident specifying probabilities in a clinical scenario in which they have significant experience.

The 95\% probability intervals from the implied distributions for the two conditions show significant overlap (Table \ref{PI}) and the 95\% probability intervals for the relative risk and odds ratio comparing the scenarios were $(0.518,2.549)$ and $(0.498,2.630)$ respectively, both containing one. Based on these assessments, the modelling does not provide evidence that hospital admissions as a result of true exacerbations would decrease on adoption of the test. 

{\bf Query 2} The top right plot in Figure \ref{dens} provides the densities of the beta distributions under the no test and test scenarios for the output quantity of query 2, the probability of missing an exacerbation when a patient suspects an exacerbation. We see only a very small overlap between the implied distributions of the proportion of patients who will experience a missed exacerbation in the absence or the presence of the test. The adoption of the test is very likely to reduce the proportion of patients missing exacerbations based on this plot.

The 95\% probability intervals (Table \ref{PI}) for test adoption and no test adoption do not overlap, supporting the inference from the implied distributions. Similarly, the 95\% probability intervals for the relative risk and the odds ratio are $(1.558,6.089)$ and $(1.689,7.439)$ respectively, with neither containing one, providing further evidence of a difference in the proportions.    

{\bf Query 3} The bottom left plot in Figure \ref{dens} provides the densities of the beta distributions under the no test and test scenarios for the output quantity of query 3, the probability of an unnecessary visit to A\&E when a patient suspects an exacerbation. We see essentially no overlap between the implied distributions for the proportions of unnecessary visits under adoption and non-adoption of the test, giving evidence that the test is highly likely to reduce unnecessary A\&E visits, even after taking into account our uncertainty.

Again the 95\% probability intervals for the implied distributions under ``test`` and ``no test`` support this conclusion, as they do not overlap (Table \ref{PI}), and so do the 95\% probability intervals for the relative risk and the odds ratio, which were $(2.708,44.305)$ and $(2.747,45.896)$ respectively, neither including one. 

\subsection{Sensitivity analysis}\label{sens}

Based on the analysis in the previous section, further investment in test development and evaluation seems promising. A sensitivity analysis was therefore deemed useful to identify key uncertainties which it would be helpful to reduce in the next phase of the evaluation. To do this, we consider the Pearson correlation between the outcome probability for each query, and each of the input probabilities that query probability depends on. The input probabilities with the highest correlation with a query probability have the strongest influence on the query outcome.

The use of correlations to assess the sensitivity of the query probabilities to the input probabilities is relatively simplistic, and more complex approaches such as the expected value of information in a value of information (VOI) analysis \citep{Jal18,Str14} would improve this aspect of the work. However, this would require information on costs and effects which are to be collected in the next stage of the test development. Thus a comprehensive VOI analysis will be conducted then.

We report the results of the sensitivity analysis for query 1. The equivalent outputs for queries 2 and 3 are omitted, but the main conclusions are given at the end of the section, and the output can be found in \citet{Gra17}.

\begin{figure}[ht]
	\centering
	\includegraphics[height=4in]{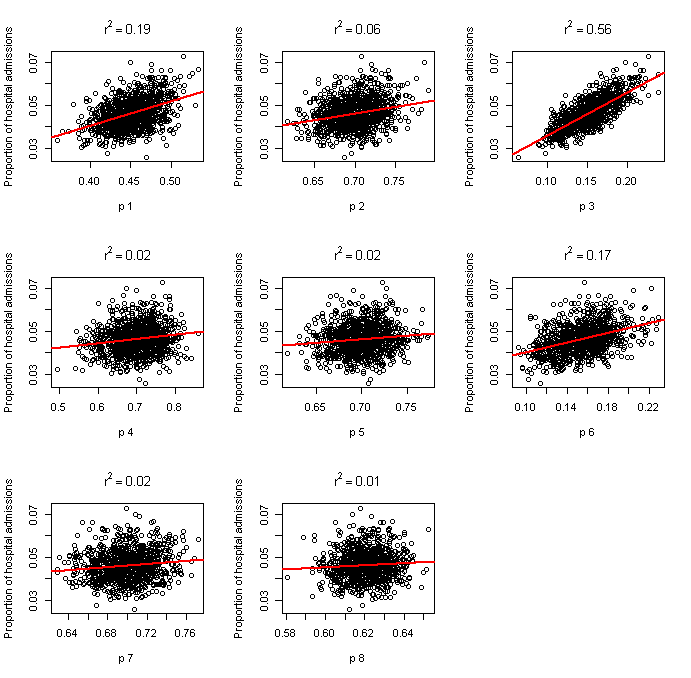}
	\caption{The proportion of hospital admissions plotted against the simulated values of each of the input probabilities for query 1 from the Monte Carlo sample, for the no test scenario.}
	\label{scatter1}
\end{figure}

\begin{figure}[ht]
	\centering
	\includegraphics[height=4in]{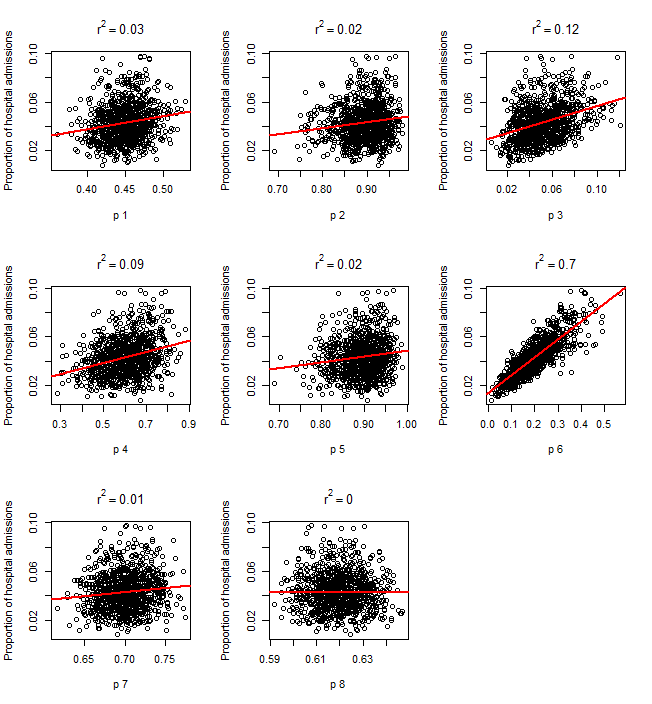}
	\caption{The proportion of hospital admissions plotted against the simulated values of each of the input probabilities for query 1 from the Monte Carlo sample, for the test adoption scenario.}
	\label{scatter2}
\end{figure}

Figures \ref{scatter1} and \ref{scatter2} show scatterplots of the proportion of hospital admissions against the simulated values of each of the input probabilities for query 1 from the Monte Carlo sample, from the no test and adoption of the test scenarios respectively. The input probabilities are defined in Table \ref{def}.

\begin{table}[ht]
	\centering {\small
		\begin{tabular}{|c|l|} \hline
			\multicolumn{2}{|c|}{Query 1: Proportion of hospital admissions} \\ \hline
			Probability & Probability definition \\ \hline
			$p_1$ & Probability of having a true exacerbation when early symptoms are present \\ 
			$p_2$ & Probability that a patient correctly diagnoses a true exacerbation \\ 
			$p_3$ & Probability that a patient refers to SC after self-diagnosis of an exacerbation \\ 
			$p_4$ & Probability that a patient refers to PC after self-diagnosis of an exacerbation \\ 
			$p_5$ & Probability that a PC clinician correctly diagnoses a true exacerbation \\ 
			$p_6$ & Probability that a PC clinician refers a patient to SC after an exacerbation diagnosis \\ 
			$p_7$ & Probability that a SC clinician admits a patient with a true exacerbatio, if referred by PC \\ 
			$p_8$ & Probability that a SC clinician admits a patient with a true exacerbation if self-referred \\ \hline
	\end{tabular} }
	\caption{Definitions of the probabilities in query 1. PC is primary care and SC is secondary care.}
	\label{def}
\end{table}


The most influential probability, with an $R^2=0.56$, in the no test scenario was $p_3$, the probability that patients refer to A\&E when suspecting an exacerbation. The probabilities $p_1$ and $p_6$ also had some impact on the distribution of the query probability ($R^2=0.19$ and $R^2=0.17$ respectively). These are the probability that early worsening symptoms evolve into an exacerbation and the probability that a healthcare professional in primary care refers a patient with a suspected exacerbation to hospital.

In the situation where the test was adopted, the most influential probabilities were $p_6$ with $R^2=0.7$ and, to a much smaller extent, $p_3$, with $R^2=0.12$. Perhaps surprisingly, the test sensitivity $p_2$, the probability that the test recognises a true exacerbation, was not strongly correlated with the probability of hospital admission. 

For query 2, the probability of having an exacerbation when some early symptoms appear (the prevalence) and the probability that patients do not recognise an exacerbation were the most influential in the no test scenario and in addition the probability that a primary clinician does not recognise a true exacerbation was influential in the test adoption scenario. These latter two probabilities both relate to the sensitivity of the diagnosis. 

For query 3, the probability that a patient self-refers to A\&E on suspicion of an exacerbation and the probability that a patient does not diagnose a true exacerbation were most influential. This latter probability is again related to the sensitivity of the diagnosis, indicating that the sensitivity of the test will be a crucial component.

\section{Conclusions and discussion}\label{conc}

In this paper, we discuss an exhaustive Bayesian approach in the evaluation of a clinical test in the early phases of development. We took the perspective of the funder/the test developer who has to decide between investing resources on the next phase of test development or allocating resources elsewhere. This decision depends on the future budget holder's decision between adopting the new clinical test or maintaining the current care pathway. We exploited methods developed within Bayesian analysis and decision theory to infer the structure of an influence diagram representing the care pathway, elicit and specify the probabilities to populate and solve the influence diagram and perform uncertainty and sensitivity analyses on the results.

For statisticians not working in early phase clinical test development, this case study illustrates some interesting challenges. The first is the need to take into account substantial uncertainty in the decision making with regards to whether to progress a potential test to more expensive development phases and what role the test would most usefully fulfill (diagnosis, monitoring, screening, etc.). The challenge here is that, while some partial data may be available on the likely performance (sensitivity, specificity, etc.) of the test in ideal lab-based conditions, and on the behaviour of the patients in the current care pathway, there is not the comprehensive data necessary for complex statistical modelling, particularly in the scenario of test adoption. Thus there is an opportunity for subjective Bayesian modelling, based on expert judgements and taking into account the available data, for example in an evidence dossier, to provide decision support incorporating current uncertainty for companies and funders. In our experience, these decision makers do make use of this decision support, provided it takes the form of a careful, structured approach such as that described in this paper. There are clear parallels with probabilistic risk assessment and structural reliability modelling in engineering in the problem structure and available data.

A second interesting aspect of early clinical modelling is the standard use of care pathway analysis in this area. This provides an informal graphical structure from which to construct a graphical model. In this paper the model used was an influence diagram. However, with other care pathways a range of graphical models could be used, such as chain event graphs, dynamic Bayesian networks, multi-regression dynamic models, and flow graphs \citep{Wil18b}. This provides a rich area of potential research for those interested in graphical modelling. There will be similar informal graphical structures in other disciplines which could be utilised in similar ways.

A third aspect of interest to statisticians is the link between this work and health economics. An early economic analysis is typical in the development of clinical tests. The graphical model developed in this paper could provide an initial structure for a health economic analysis, by incorporating the costs, as well as the benefits, of the addition of the test into a value node in the influence diagram. This could provide a basis, for example, for a value of information analysis \citep{Jal18,Str14}.

Based on the analyses we carried out, we concluded that, if the diagnostic test was adopted, we have not found any evidence that hospital admissions for true COPD exacerbations would decrease, but we did find evidence that the number of missed exacerbations would reduce and we also found evidence that the number of unnecessary A\&E visits from COPD patients would reduce. Further, based on the sensitivity analysis, the most influential variables were found to be the probability that a patient would present at A\&E with a suspected exacerbation and the sensitivity of the test. These findings will inform the appropriate choice of the target population and early adopters, and guide the refinement of the diagnostic test. These conclusions depend on the model assumptions, especially on the structural assumptions and on the proposed role of the test. 

The modelling carried out in this paper has been static, in the sense that we have not modelled the progression of the disease. The use of dynamic Bayesian networks or continuous time Bayesian networks could extend the approach to incorporate this temporal progression \citep{Nod02,Dea90}. 

The next stage of development is to finalise the device and test it in a patient population. This will allow us to update the influence diagram with the new evidence collected, and assess the structure of the model and inferences which we have made from it. In particular, with observations on some of the model inputs, we would expect the uncertainty in the query distributions to reduce, providing a clearer message, particularly for query 1.

\bibliographystyle{plainnat}
\bibliography{bib/MDMv1}

\section*{Acknowledgement}

	The authors would like to thank Sue Hart from the Academic Health Sciences Network (AHSN) for co-ordinating the contact with other experts and taking part as an expert in the elicitations, the AHSN for their support and Joy Allen for supporting the data collection and advice.
	
	This research was supported by the NIHR MedTech In vitro diagnostic Co-operatives scheme (ref MIC-2016-014). The views expressed are those of the authors and not necessarily those of the NHS, the NIHR or the Department of Health.

\end{document}